\documentclass{ws-procs975x65}

\begin{document}

\title{Theoretical interpretation of ``long'' and ``short'' GRBs}

\author{C.L. Bianco,$^{1,2}$ M.G. Bernardini,$^{1,2}$ L.Caito,$^{1,2}$ P. Chardonnet,$^{1,4}$ M.G. Dainotti,$^{1,2}$ F. Fraschetti,$^5$ R. Guida,$^{1,2}$ R. Ruffini,$^{1,2,3}$ S.-S. Xue$^{1}$}

\address{
$^1$ ICRANet and ICRA, Piazzale della Repubblica 10, I-65122 Pescara, Italy.\\
$^2$ Dip. di Fisica, Universit\`a di Roma ``La Sapienza'', Piazzale Aldo Moro 5, I-00185 Roma, Italy.\\
$^3$ ICRANet, Universit\'e de Nice Sophia Antipolis, Grand Ch\^ateau, BP 2135, 28, avenue de Valrose, 06103 NICE CEDEX 2, France.\\
$^4$ Universit\'e de Savoie, LAPTH - LAPP, BP 110, F-74941 Annecy-le-Vieux Cedex, France.\\
$^5$ CEA/Saclay, F-91191 Gif-sur-Yvette Cedex, Saclay, France.\\
E-mails: bianco@icra.it, maria.bernardini@icra.it, letizia.caito@icra.it, chardon@lapp.in2p3.fr, dainotti@icra.it, fraschetti@icra.it, roberto.guida@icra.it, ruffini@icra.it, xue@icra.it.
}

\begin{abstract}
Within the ``fireshell'' model we define a ``canonical GRB'' light curve with two sharply different components: the Proper-GRB (P-GRB), emitted when the optically thick fireshell of electron-positron plasma originating the phenomenon reaches transparency, and the afterglow, emitted due to the collision between the remaining optically thin fireshell and the CircumBurst Medium (CBM). We here present the consequences of such a scenario on the theoretical interpretation of the nature of ``long'' and ``short'' GRBs.
\end{abstract}

\bodymatter

\section{Introduction}

We assume that all GRBs, both ``long'' and ``short'', originate from the gravitational collapse to a black hole.\cite{2001ApJ...555L.113R,2007AIPC..910...55R} The $e^\pm$ plasma created in the process of the black hole formation expands as an optically thick and spherically symmetric ``fireshell'' with a constant width in the laboratory frame, i.e. the frame in which the black hole is at rest.\cite{1999A&A...350..334R} We have only two free parameters characterizing the source: the total energy of the $e^\pm$ plasma $E_{e^\pm}^{tot}$ and the $e^\pm$ plasma baryon loading $B\equiv M_Bc^2/E_{e^\pm}^{tot}$, where $M_B$ is the total baryons' mass.\cite{2000A&A...359..855R} These two parameters fully determine the optically thick acceleration phase of the fireshell, which lasts until the transparency condition is reached and the Proper-GRB (P-GRB) is emitted.\cite{2001ApJ...555L.113R,2007AIPC..910...55R} The afterglow emission then starts due to the collision between the remaining optically thin fireshell and the CircumBurst Medium (CBM).\cite{2001ApJ...555L.113R,2007AIPC..910...55R,2004ApJ...605L...1B,2005ApJ...620L..23B,2005ApJ...633L..13B} It clearly depends on the parameters describing the effective CBM distribution: its density $n_{cbm}$ and the ratio ${\cal R}\equiv A_{eff}/A_{vis}$ between the effective emitting area of the fireshell $A_{eff}$ and its total visible area $A_{vis}$.\cite{2002ApJ...581L..19R,2004IJMPD..13..843R,2005IJMPD..14...97R,2007A&A...471L..29D}

\begin{figure}
\includegraphics[width=0.495\hsize,clip]{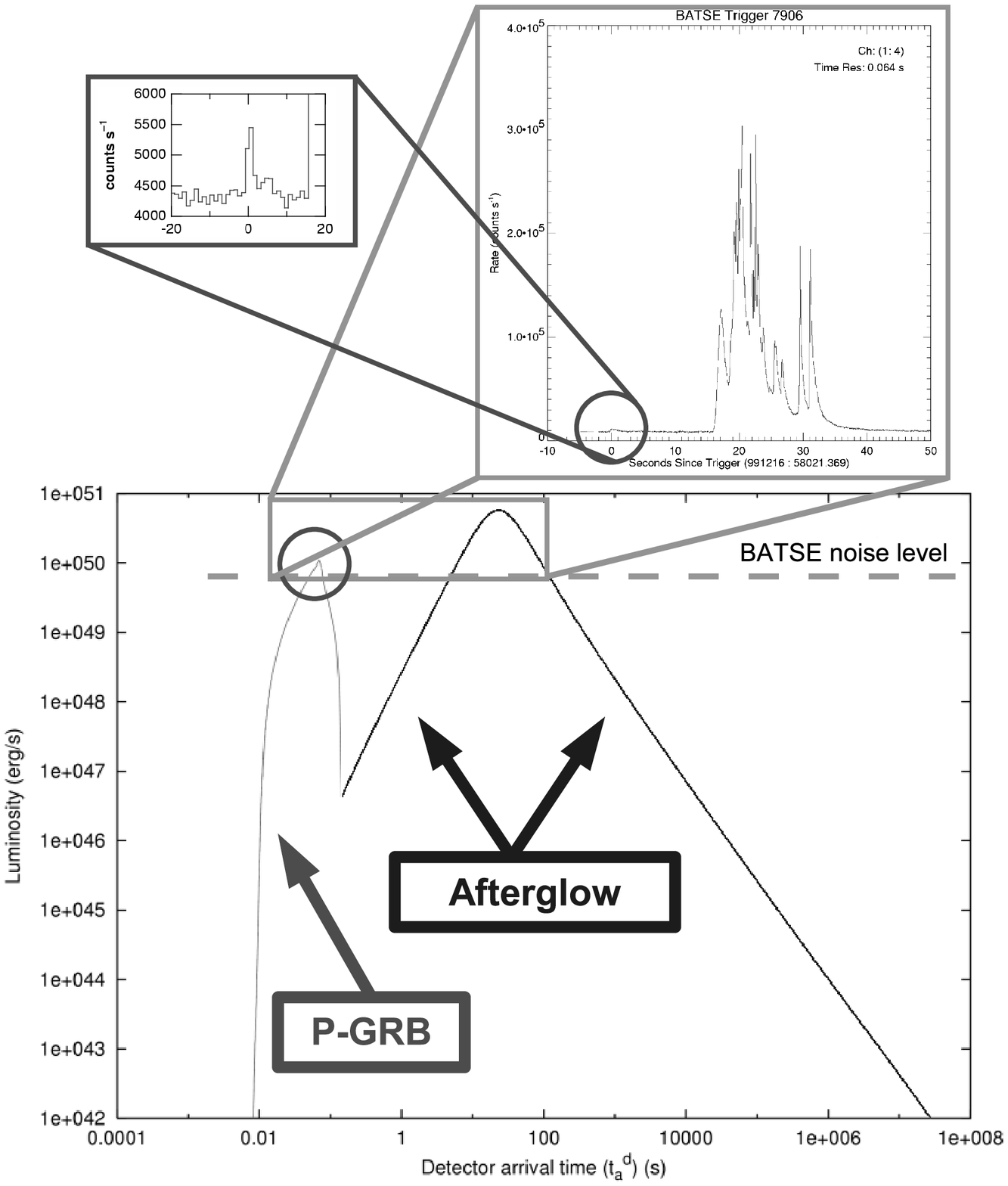}
\includegraphics[width=0.495\hsize,clip]{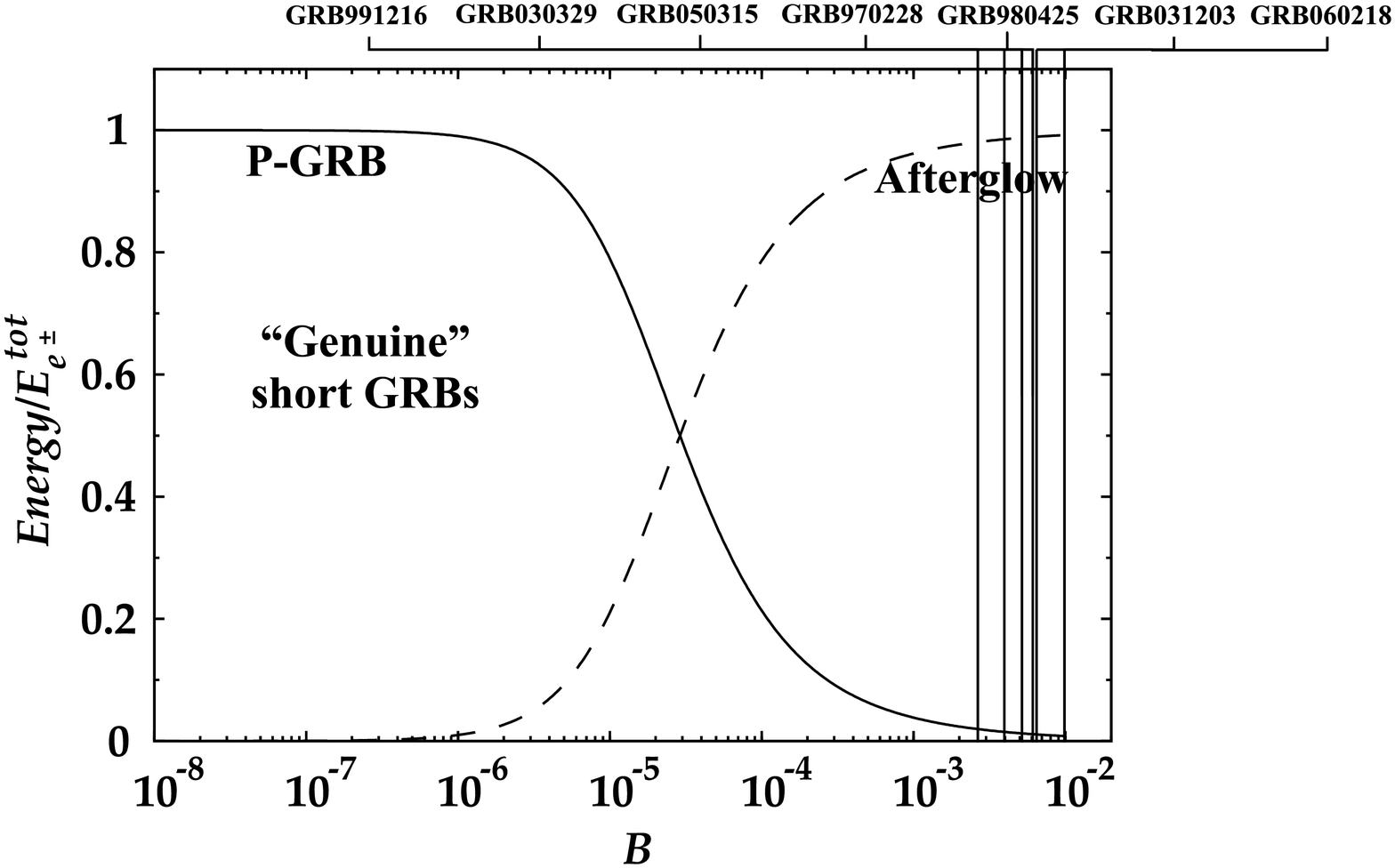}
\caption{{\bf Left:} The ``canonical GRB'' light curve theoretically computed for GRB 991216. The prompt emission observed by BATSE is identified with the peak of the afterglow, while the small precursor is identified with the P-GRB. For this source we have $B\simeq 3.0\times 10^{-3}$.\cite{2001ApJ...555L.113R,2002ApJ...581L..19R,2003AIPC..668...16R,2005AIPC..782...42R} {\bf Right:} The energy radiated in the P-GRB (solid line) and in the afterglow (dashed line), in units of the total energy of the plasma ($E_{e^\pm}^{tot}$), are plotted as functions of the $B$ parameter. Also represented are the values of the $B$ parameter computed for GRB 991216, GRB 030329, GRB 980425, GRB 970228, GRB 050315, GRB 031203, GRB 060218. Remarkably, they are consistently smaller than, or equal to in the special case of GRB 060218, the absolute upper limit $B \lesssim 10^{-2}$.\cite{2000A&A...359..855R} The ``genuine'' short GRBs have a P-GRB predominant over the afterglow: they occur for $B \lesssim 10^{-5}$.\cite{2001ApJ...555L.113R,2007A&A...474L..13B}}
\label{figx}
\end{figure}

\section{The ``canonical GRB'' scenario}

Unlike treatments in the current literature,\cite{2005RvMP...76.1143P,2006RPPh...69.2259M} we define a ``canonical GRB'' light curve with two sharply different components (see Fig.~\ref{figx} and Refs.~\refcite{2001ApJ...555L.113R,2007AIPC..910...55R,2007A&A...474L..13B,2007A&A...471L..29D,bianco_ita-sino}):
\begin{itemize}
\item \textbf{The P-GRB}, which has the imprint of the black hole formation, an harder spectrum and no spectral lag;\cite{2001A&A...368..377B,2005IJMPD..14..131R}
\item \textbf{the afterglow}, which presents a clear hard-to-soft behavior;\cite{2004IJMPD..13..843R,2005ApJ...634L..29B,2006ApJ...645L.109R} the peak of the afterglow contributes to what is usually called the ``prompt emission''.\cite{2001ApJ...555L.113R,2007A&A...471L..29D,2006ApJ...645L.109R}.
\end{itemize}
The ratio between the total time-integrated luminosity of the P-GRB (namely, its total energy) and the corresponding one of the afterglow is the crucial quantity for the identification of GRBs' nature. Such a ratio, as well as the temporal separation between the corresponding peaks, is a function of the $B$ parameter (see Fig.~\ref{figx} and Ref.~\refcite{2001ApJ...555L.113R}).

When $B \lesssim 10^{-5}$, the P-GRB is the leading contribution to the emission and the afterglow is negligible: we have a ``genuine'' short GRB.\cite{2001ApJ...555L.113R} When $10^{-4} \lesssim B \lesssim 10^{-2}$, instead, the afterglow contribution is generally predominant. Still, this case presents two distinct possibilities: the afterglow peak luminosity can be either \textbf{larger} or \textbf{smaller} than the P-GRB one.\cite{2007A&A...474L..13B,bianco_ita-sino} The simultaneous occurrence of an afterglow with total time-integrated luminosity larger than the P-GRB one, but with a smaller peak luminosity, can indeed be explained in terms of a peculiarly small average value of the CBM density ($n_{cbm} \sim 10^{-3}$ particles/cm$^3$), compatible with a galactic halo environment (``fake'' short GRBs).\cite{2007A&A...474L..13B,bianco_ita-sino}

\end{document}